\documentclass[referee]{mn2e}

\usepackage{graphicx}
\usepackage{amsmath}

\title[On the gamma-ray variability from the Crab Nebula]
{On the variability of the GeV and multi-TeV gamma-ray emission from the Crab Nebula}
\author[W. Bednarek \& W. Idec]
{W. Bednarek \& W. Idec\\
Department of Astrophysics, University of \L \'od\'z,
ul. Pomorska 149/153, 90-236 \L \'od\'z, Poland \\
bednar@astro.phys.uni.lodz.pl; idec@kfd2.phys.uni.lodz.pl}
\begin{document}

\date{Accepted . Received ; in original form }

\pagerange{\pageref{firstpage}--\pageref{lastpage}} \pubyear{2007}

\maketitle

\label{firstpage}

\begin{abstract}
Recently the AGILE $\gamma$-ray telescope has reported the enhanced $\gamma$-ray emission above 100 MeV from the direction of the Crab Nebula during a period of a few days. This intriguing observation has been confirmed by the Fermi-LAT telescope. This emission does not show evidences of pulsations with the Crab pulsar. It seems that it originates at the shock region created as a result of the interaction of the pulsar wind with the nebula. 
We propose that such variable $\gamma$-ray emission originate in the region behind the shock when the electrons can be accelerated as a result of the reconnection of the magnetic field compressed by the decelerating pulsar wind.
The natural consequence of such interpretation is the prediction that the Crab Nebula $\gamma$-ray spectrum produced by electrons as a result of the inverse Compton scattering of soft radiation to multi-TeV energies should also show synchronous variability on the time scales as observed at GeV energies by the AGILE and Fermi-LAT telescopes. We calculate how the end of the IC component of the Crab Nebula $\gamma$-ray spectrum should look like during the quiescent and the flare GeV $\gamma$-ray emission. We conclude that the variability of the multi-TeV $\gamma$-ray spectrum from the Crab Nebula might in principle be responsible for the differences between the spectral features reported by the HEGRA and HESS Collaborations
at the multi-TeV energies.
\end{abstract}
\begin{keywords} neutron stars: nebulae --- individual: Crab Nebula --- radiation mechanisms: non-thermal --- gamma-rays: theory  
\end{keywords}

\section{Introduction}

The Crab Nebula $\gamma$-ray emission has been established as a standard candle for the $\gamma$-ray astronomy (e.g. Meyer et al.~2010). This conclusion based mainly on the observations at energy ranges between $\sim 100$ MeV and a few GeV (satellite telescopes) and between
$\sim 100$ GeV and $\sim 10$ TeV (Cherenkov telescopes). The theoretical interpretation of the emission from the Crab Nebula suggests that this may not be exactly the case.
The $\gamma$-ray spectrum is widely interpreted in the two component radiation model in which the lower energy emission is due to the synchrotron process and the higher energy emission results as a consequence of the Inverse Compton (IC) process. It has been argued that
the ends of the synchrotron and IC components can flicker due to the non-stationary acceleration of leptons at the pulsar wind shock. First evidences of the variability of the end of synchrotron component has been reported based on the analysis of the EGRET data
(de Jager et al.~1996, Ramanamurthy et al.~1995). Also situation has not been clear at the highest observer energies (above $\sim 10$ TeV), where some measurements seemed to  be contradictory (Aharonian et al.~2004, Aharonian et al.~2006). 

The spectrum from the Crab Nebula at energies 100-400 MeV is very steep, with the differential spectral index close to 4 (Abdo et al.~2010). It cuts off at about $\sim 100$ MeV, i.e. at clearly larger energies than previously reported $\sim 25$ MeV cut-off derived from the EGRET measurements (Kuiper et al.~2001). On the other hand, the $\gamma$-ray spectrum at GeV energies is very flat linking correctly with the TeV spectrum measured by the Cherenkov telescopes (e.g. Albert et al.~2008). Unexpectedly, the 
AGILE telescope has recently reported an enhanced $\gamma$-ray emission above $100$ MeV from the Crab Nebula by a factor of 2-3 during September 19-21, 2010 in respect to steady emission (Tavani et al.~2010, 2011). This observation has been confirmed by the Fermi-LAT telescope which observed enhanced $\gamma$-ray emission 
during the interval September 18-22, 2010 (Buehler et al.~2010, Abdo et al.~2011). The $\gamma$-ray flux reached the value of $(606\pm 43)\times 10^{-8}$ ph. cm$^{-2}$ s$^{-1}$ above 100 MeV.
The flaring component has a differential spectral index $2.49\pm 0.14$ and is coincident with the Crab Nebula. 
The flare consists from three separated enhancements lasting 6-12 hours
(Balbo et al.~2010).
The lack of pulsed emission suggests that the flare is probably related to the Crab Nebula rather than the pulsar (Hays et al.~2010).

The enhanced $\gamma$-ray emission at about 1 TeV has been also reported by 
the ARGO-YBJ air shower array (Aielli et al.~2010). This flaring component extends up to September 27th 2010, i.e. it lasts a few days longer than the 100 MeV emission reported by the satellite telescopes. The flux measured during the time interval September 17-22nd, 2010 was about 3-4 time higher than usual. Unfortunately, this enhanced emission has not been confirmed by the observations of the MAGIC (Mariotti et al.~2010) and VERITAS Collaborations (Ong et al.~2010).

The increased optical emission about 3 arcsec east of the Crab pulsar and from the wisp north-west of the pulsar has been reported by the Hubble Space Telescope
(Caraveo et al.~2010). This corresponds to the brightening of this same region in the X-rays as reported by the Chandra observations (Tennant et al.~2010).
However, other X-ray telescopes have not reported any significant change in the Crab Nebula spectrum and morphology (Sakamoto et al.~2010, Shaposhnikov et al.~2010).
Moreover, no any glitch of the Crab pulsar has been noted during the last 60 days before the $\gamma$-ray flare (Espinoza et al.~2010).

In this paper we investigate the hypothesis that the variable $\gamma$-ray emission
above $100$ MeV is due to the synchrotron radiation from electrons which are accelerated near the pulsar wind shock region to different maximum energies. 
We expect that variable GeV $\gamma$-ray emission should be accompanied by the variable $\gamma$-ray emission above a few TeV from the Inverse Compton process.
Discovery of such variable emission by the present Cherenkov telescopes (HESS, MAGIC, VERITAS) and the future Cherenkov Telescope Array (CTA) will allow to constrain the physical processes at the acceleration site and investigate the connections between the pulsar and the nebula. Note that recently 
Komissarov \& Lyutikov~(2010) proposed the localization of the $\gamma$-ray flare emission region with the compact inner knot, based on their recent results on relativistic MHD simulations of the Crab nebula.

\section{A scenario for variable emission from the Crab Nebula}

The variability time scale of the $\gamma$-ray emission on the level of days can be understood assuming that the emission comes only from a part of the pulsar wind shock. Therefore, either only a small region of the axially symmetric pulsar wind shock is excited or the emission region is moving relativistically towards the observer. 
The axial symmetry of the pulsar wind (with the axis parallel to the rotational axis of the pulsar) is consistent with the observations of the jet-torus morphology of the Crab Nebula (e.g. Hester 2008) and with the MHD simulations of the pulsar wind structure (Komissarov \& Lyubarsky~2003).
The first hypothesis seems unlikely since the wisp regions have already extensions which are larger than the light travel distance scale corresponding to 3-4 day time scale variability. Here we consider the second hypothesis in which the $\gamma$-ray emission region is related to the shock region which still moves with substantial Lorentz factor, $\gamma_{\rm sh}$, in the outward direction from the pulsar. This might be the region in the pulsar wind which is in the process of being decelerated at the pulsar wind shock region.
It is expected that in this region efficient reconnection process of the pulsar wind magnetic field can occur. As a result, good conditions are produced for acceleration of particles to the highest possible energies.
The first wisps in the Crab Nebula appear at the distance of $R_{\rm sh}\sim 7\times 10^{16}$ cm from the pulsar (Caraveo et al.~2010). In fact, recent optical observations of the wisp confirm its variability which may be related to the observed $\gamma$-ray variability by the AGILE and Fermi.

The observed day time scale $\gamma$-ray variability   
allows to constrain $\gamma_{\rm sh}$. We assume that emission extends along the axially symmetric shock structure as observed in the Crab Nebula and expected from the MHD simulations (Komissarov \& Lyubarsky 2003). The turbulence which arrive to the shock is relatively thin, i.e. it is thinner than the light crossing time scale multiplied by the Lorentz factor of the emission region. The distant observer is able to detect emission only from a part of the shock with the opening angle $\alpha$ (see Fig.~1). Based on such scenario we can estimate $\sin\alpha$ on,
\begin{eqnarray}
\sin\alpha\approx
\sqrt{(c\tau_{\rm v}/R_{\rm sh})^2 + 2(c\tau_{\rm v}/R_{\rm sh})}.
\label{eq1}
\end{eqnarray}
\noindent
Produced radiation is collimated within the angle $\alpha$ for the Lorentz factor of the shock 
of the order of,
\begin{eqnarray}
\gamma_{\rm sh}\sim 1/\alpha\approx \sqrt{R_{\rm sh}/(2c\tau_{\rm v})}, 
\label{eq2}
\end{eqnarray}
\noindent
provided that the variability time scale of the $\gamma$-ray emission is $\tau_{\rm v} << R_{\rm sh}/c$. In the case of the Crab Nebula shock with dimension $R_{\rm sh} = 7\times 10^{16}$ cm, the minimum Lorentz factor of the emission region should be of the order of $\gamma_{\rm sh}\approx 3.7\tau_{\rm d}^{-1/2}$ where $\tau_{\rm v} = 1\tau_{\rm d}$ days. We conclude that for the sub-day time scale variability (as observed in $\gamma$-rays by the Fermi-LAT telescope, Balbo et al.~2010), the Lorentz factor of the emission region in the shock has to be at least mildly relativistic. We suggest that the variable $\gamma$-ray emission observed from the Crab Nebula comes from a part of the pulsar wind which is just behind the shock and still moves with relativistic velocities.

\begin{figure}
\vskip 5.2truecm
\includegraphics{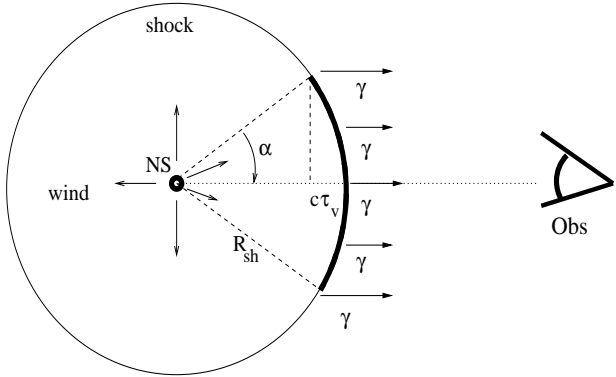}
\caption{Schematic representation of the Crab Nebula. The pulsar (NS) produce relativistic axially symmetric wind which creates a shock at the distance $R_{\rm sh}$. $\gamma$-rays are produced behind the shock in the region of strong deceleration of the pulsar wind. Only those produced close to the part of the shock (thick part defined by the angle $\alpha$) can reach the observer (Obs). However, they reach the observer at different time due to the curvature of the shock structure. This time difference is limited by the observed variability time scale of the $\gamma$-ray emission ($c\tau_{\rm v}$).
It can be estimated for known $R_{\rm sh}$ and $\tau_{\rm v}$. The angle $\alpha$ is related to the collimation of the $\gamma$-ray emission which is caused by the relativistic motion of the emission region with the Lorentz factor $\gamma_{\rm sh}$.}
\label{fig1}
\end{figure}

We constrain the physical parameters of the emission region in the fluid frame (in which the electric field vanishes) based on the observations. This reference frame moves towards the observer with the Lorentz factor $\gamma_{\rm f}$. 
We consider relativistic electrons with characteristic Lorentz factors $\gamma_{\rm br}^{\rm f}$ in the fluid reference frame which are immersed in the perpendicular magnetic field with the strength in the fluid frame $B_{\rm f}$. 
The synchrotron spectrum from the Crab Nebula extends up to a hundred MeV, showing a break at, $\varepsilon_{\rm obs}\sim$100 MeV, during the quiescent stage (Abdo et al.~2010). This break can be related to a break in the electron spectrum according to,
\begin{eqnarray}
\varepsilon^{\rm syn}_{\rm br}\approx m_{\rm e}(B_{\rm f}/B_{\rm cr})(\gamma_{\rm br}^{\rm f})^2\approx \varepsilon_{\rm obs}/\gamma_{\rm f},
\label{eq3}
\end{eqnarray}
\noindent
where the critical magnetic field is $B_{\rm cr}\approx 4.4\times 10^{13}$ G, and $m_{\rm e}$ is the electron rest mass.
Applying the above value for the break in the synchrotron spectrum, $\varepsilon_{\rm obs}$, we get the constraint on the product of the magnetic field strength within the emission region and the Lorentz factor of radiating electrons,
\begin{eqnarray}
B_{\rm f}(\gamma_{\rm br}^{\rm f})^2\approx 8.8\times 10^{15}/\gamma_{\rm f}~~~{\rm G}. 
\label{eq4}
\end{eqnarray}
On the other hand, the $\gamma$-ray flare observed by the AGILE  and the Fermi telescopes lasted for a few days and show a clear variability on a sub-day time scale, $\tau_{\rm obs}$ (Balbo et al.~2010). The duration of the flare has to be comparable (or longer) to the cooling time scale of electrons on the synchrotron process in the fluid frame, $\tau_{\rm syn}^{\rm cool}\approx \tau_{\rm obs}\gamma_{\rm f}$. 
By comparing these two time scales, we get the lower limit on,
\begin{eqnarray}
B_{\rm f}^2\gamma_{\rm br}^{\rm f}\approx 6.5\times 10^3/(\tau_{\rm d}\gamma_{\rm f})~~~{\rm G^2}.
\label{eq5}
\end{eqnarray}
\noindent
These two conditions allows us to estimate the magnetic field strength at the acceleration region in the fluid frame on,
\begin{eqnarray}
B_{\rm f} > 1.7\times 10^{-3}\tau_{\rm d}^{-2/3}\gamma_{\rm f}^{-1/3}~~~{\rm G}.
\label{eq6}
\end{eqnarray}
\noindent
The magnetic field strength in the observer's frame is, $B_{\rm obs} = B_{\rm f}\gamma_{\rm f}$. For the variability time scale observed by the Fermi (of the order of 0.5 day, Balbo et al.~2010), we obtain the lower limit on $B_{\rm f}\approx 2.7\times 10^{-3}\gamma_{\rm f}^{-1/3}\approx 1.6\times 10^{-3}$ G, assuming that $\gamma_{\rm f} = \gamma_{\rm sh}$. The magnetic field as seen  from the observer reference frame is estimated on $B_{\rm obs}\approx 8.4\times 10^{-3}$ G.
On the other hand, a simple extrapolation of the magnetic field from the Crab pulsar surface up to the location of the first optical wisps in the Crab Nebula allows to estimate their magnetic field strength. We assume a dipole magnetic field structure in the pulsar inner magnetosphere below the light cylinder radius, $B(R_{\rm LC}) = B_{\rm NS}(R_{\rm NS}/R_{\rm LC})^3$, and a toroidal structure in the pulsar wind, $B(R) = B(R_{\rm LC})(R_{\rm LC}/R)$, where the surface magnetic field of the Crab pulsar $B_{\rm NS} = 6\times 10^{12}$ G, the light cylinder radius $R_{\rm LC} = cP_{\rm NS}/(2\pi)\approx 1.6\times 10^8$ cm, $B(R_{\rm LC})\approx 1.5\times 10^6$ G is the magnetic field strength at the light cylinder, the Crab pulsar rotational period is $P_{\rm NS} = 33$ ms, and $c$ is the velocity of light. The distance of the optical wisp from the pulsar is $R_{\rm wisp}\approx 7\times 10^{16}$ cm, estimated from the angular distance of 3 arcsec (Caraveo et al.~2010), and for the distance to the Crab Nebula of 1650 pc. This wisp has been recently reported as showing optical variability. Then, the magnetic field strength at the distance of the wisp is $B_{\rm wisp} = B(R_{\rm LC})(R_{\rm LC}/R_{\rm wisp})\sim 3.5\times 10^{-3}$ G. This estimate of the magnetic field fits well to the above estimate based on the variability time scale of the $\gamma$-ray emission.

In any way, this lower limit is significantly larger than the magnetic field strength within the whole volume of the Crab Nebula estimated on $\sim 10^{-4}$ G. Such low values of the magnetic field has been already predicted based on the spherically symmetric MHD model developed by Kennel \& Coroniti (1984). They have been successfully applied in modelling of the multiwavelength non-thermal emission from the Crab Nebula in terms of the popular Synchrotron self-Compton model (see e.g., de Jager \& Harding~1992, Atoyan \& Aharonian~1995, or recently Meyer, Horns \& Zechlin~2010). 
The discrepancy between the magnetic field values estimated for the inner wisp and the whole Crab Nebula can be naturally understood in terms of the recent
axially symmetric MHD model discussed by Komissarov \& Lyutikov~(2010).
In this model the observed features in the inner part of the Crab Nebula (i.e. the inner knot and possibly also the wisp regions) are created by parts of the pulsar wind shock which lays significantly closer to the pulsar than the equatorial part of the pulsar wind shock. Therefore, the magnetic field strength at the considered by us the wisp region can be much stronger in respect to the whole nebula. The wisp region is quite extended but for the $\gamma$-ray flaring might be responsible only a part of this region which is more directed towards the observer (see Fig~1).
The possible variability of the pulsar wind termination shock (discovered in MHD simulations by Camus et al.~2009) can be responsible for the change of conditions in the acceleration region directed towards the observer at the Earth.

\section{Acceleration of electrons}

In general, the variable emission at the GeV energies observed from the Crab Nebula might be due to the change in the magnetic field strength at the acceleration site or the change of the plasma conditions which determines the acceleration process. Its influence on the acceleration of particles is usually described by the so called acceleration parameter. 
We limit our considerations only to the second case since it is difficult to imagine situation in which the magnetic field, which source lays in the pulsar, can change significantly on a time scale of a few days in the pulsar wind.

The problem is how electrons can be accelerated to considered energies in the classical acceleration processes in such a relatively strong magnetic field. In the case of the shock acceleration scenario or acceleration in a turbulent region, the maximum Lorentz factors of electrons are limited by the synchrotron energy losses to $\gamma_{\rm max}\approx 10^9(\chi_{-1}/B_{-3})^{1/2}$, 
where $\chi = 0.1\chi_{-1}$ is the acceleration coefficient, and the magnetic field strength at the acceleration site is defined in the shock reference frame. It is  scaled with $B = 10^{-3}B_{-3}$ G.
This limit has been obtained from the comparison of the energy gains of electrons from the shock acceleration mechanism, ${\dot P_{\rm acc}} = \chi E_{\rm e}/R_{\rm L}$, with their synchrotron energy losses,
${\dot P_{\rm syn}} = (4/3)\sigma_{\rm T}(B^2/8\pi)\gamma^2$,
where $E_{\rm e} = m_{\rm e}\gamma_{\rm e}$ is the energy of electrons, $R_{\rm L}$ is the Larmor radius of electrons, and $\sigma_{\rm T}$ is the Thomson cross section.

The maximum energies of synchrotron photons produced by electrons with the Lorentz factors, $\gamma_{\rm e}$, are independent on the magnetic field strength. They can be estimated from  $\varepsilon_{\rm max}\approx 11\chi_{-1}$ MeV (see Eq.~3). Note that the maximum energies of synchrotron photons produced by electrons accelerated in the shock scenario can not exceed $\sim 100$ MeV since $\chi\le 1$. This is also true in the case of relativistic shocks if considered in the shock reference frame (Kirk \& Reville~2010). However, in the case of relativistic shocks these maximum energies observed in the rest frame can be enlarged by the Doppler factor of the shock as shown in the analytical calculations by Achterberg et al.~(2001). Therefore, detection of synchrotron GeV emission from the Crab Nebula requires that the emission region moves towards the
observer with the Doppler factor of the order of $\sim$10, consistent with the lower limit estimated in Sect.~2. On the other hand, Monte Carlo simulations of the acceleration of particles by the first order Fermi shocks propagating in a realistically modelled turbulent magnetic fields show that the process may not be efficient and the maximum energies of accelerated particles can be lower than expected in previous modellings (Niemiec \& Ostrowski~2004, Niemiec, Ostrowski \& Pohl~2006).  

Another possibility, originally considered by Coroniti (1990) and Michel~(1994), is the acceleration of electrons in the reconnection regions of the magnetic field in the pulsar wind before it reaches the shock region (e.g. Lyubarsky~1996, Lyubarsky \& Kirk~2001, Kirk \& Skjaeraasen~2003).
In the case of acceleration in the reconnection regions (along the magnetic field lines), electron energy does not need to be limited by the synchrotron energy losses (Kirk 2004). Electrons are injected into the magnetic field of the pulsar wind zone producing synchrotron $\gamma$-rays with energies overcoming the synchrotron energy loss limit on the shock acceleration process mentioned above.

Due to the observed shape of the spectrum from the Crab Nebula, we assume that electrons reach the equilibrium spectrum which can be described by a differential power law spectrum with the characteristic cut-off at $\gamma_{\rm br}$,
$dN/d\gamma_{\rm e} = \gamma^{-\beta}\exp(-\gamma/\gamma_{\rm br})$,
where $\beta$ is the spectral index considered in the range 3.0-3.6 in order to be consistent with the observed synchrotron spectrum at energies below the break. 
Electrons with such equilibrium spectrum produce synchrotron photons up to GeV energies
and also TeV $\gamma$-rays by scattering variety of soft radiation fields inside or close to their acceleration site. Between them the best defined is the Microwave Background Radiation (MBR). The level of the soft synchrotron emission produced by low energy electrons fulfilling the whole Crab Nebula can not be so well defined in the region of the acceleration of electrons close to the pulsar wind shock.

\section{Gamma-rays from the Crab Nebula}

The reported variability of a hundred MeV - GeV $\gamma$-ray emission from the Crab Nebula
seems to be naturally interpreted as a result of the non-stationary acceleration of the end of the electron spectrum in the region when the pulsar wind interacts with the nebula. We propose that such emission variable on a short time scale likely comes from the region just behind the pulsar wind shock which still moves mildly relativistically. These electrons should produce synchrotron emission which sometimes can extend up to GeV energies.
On the other hand, these same electrons can produce also variable multi-TeV $\gamma$-rays as a result of the inverse Compton scattering of the MBR and very low energy (in radio range) synchrotron radiation.

\begin{figure}
\vskip 12.5truecm
\includegraphics{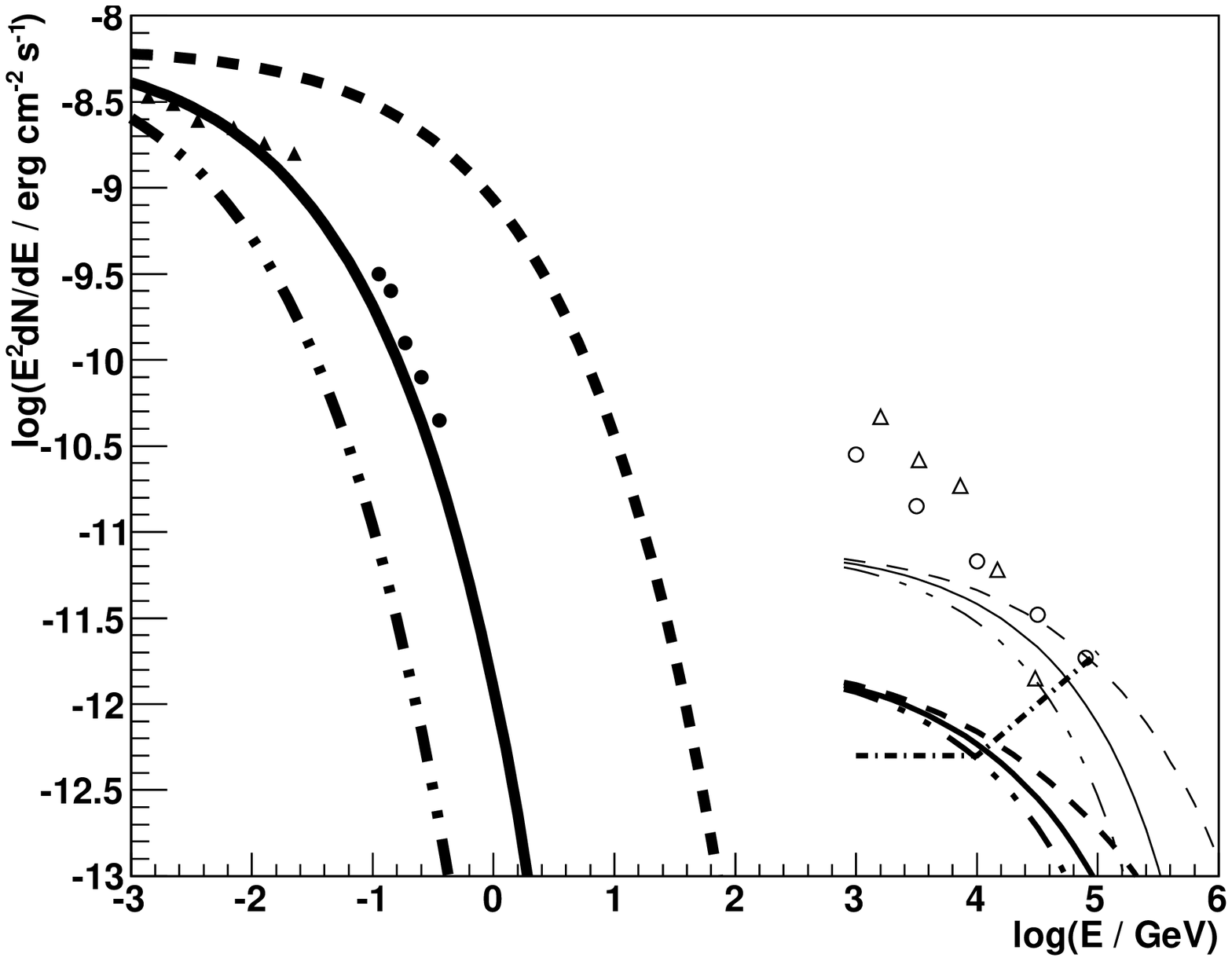}
\includegraphics{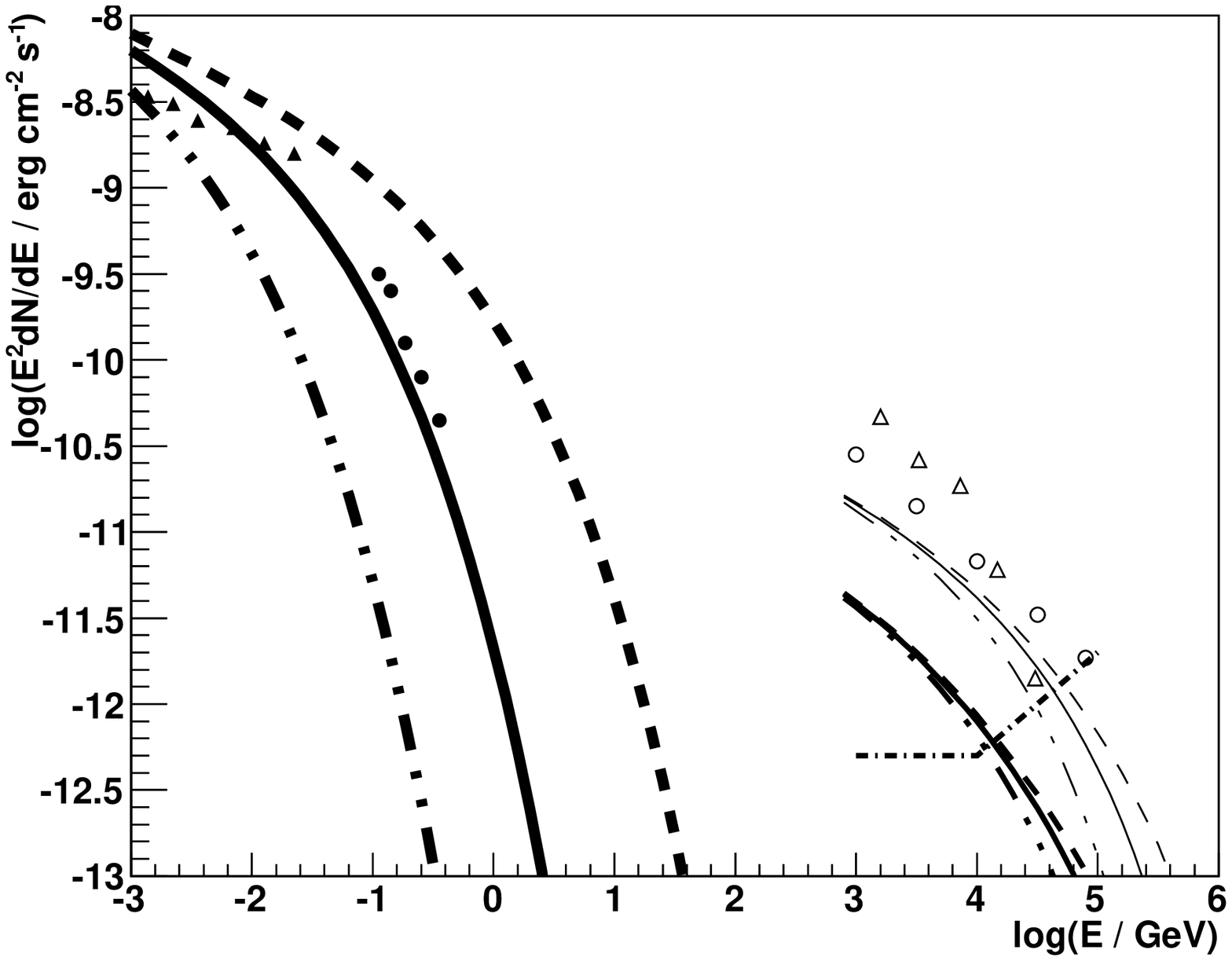}
\includegraphics{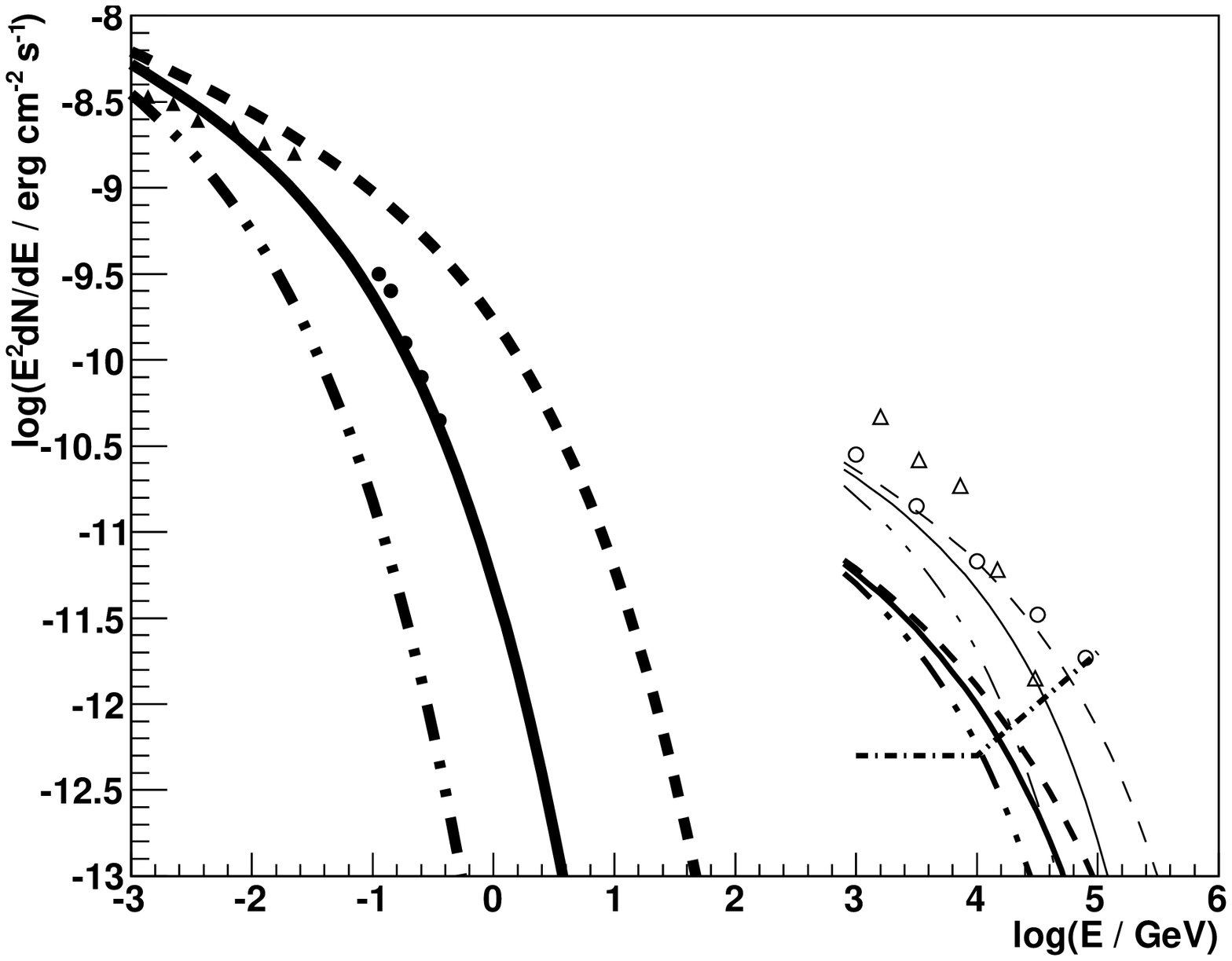}
\caption{The $\gamma$-ray spectrum from the Crab Nebula is compared with the calculations of the synchrotron spectrum (thick curves) and the IC spectra produced by electrons as a result of scattering of the MBR (thin curves) and low energy synchrotron emission within the nebula
(middle thick curves) in the quiescent, flaring, and supposed super-quiet episodes. (a) The magnetic field in the emission region is fixed on $B_{\rm sh} = 2\times 10^{-3}$ G, 
the spectral index of electron spectrum is $\beta = 3$.),
and the Lorentz factors of electrons at the break in the spectrum are fixed on $\gamma_{\rm e}^{\rm q} = 7\times 10^8$ (solid curves, for the quiescent emission), $\gamma_{\rm e}^{f} = 3\times 10^9$ (dashed curves, for the flaring emission), and $2.3\times 10^8$ (dot-dot-dashed curves).  
(b) As in figure (a) but for the spectral index of electrons $\beta = 3.6$.
(c) As in figure (b) but for $B_{\rm sh} = 4\times 10^{-2}$ G, and the Lorentz factors of electrons at the break fixed on $\gamma_{\rm e}^{\rm q} = 2\times 10^8$ (for the quiescent emission) and $\gamma_{\rm e}^{f} = 8\times 10^8$ (for the flaring emission),
and $7\times 10^7$. 
The $\gamma$-ray spectrum in the quiescent stage is measured by the satellite telescopes above 100 MeV: COMPTEL (filled triangles, Kuiper et al.~2001) and Fermi (filled circles, Abdo et al.~2010) and by the Cherenkov telescopes above 1 TeV: HEGRA (open circles, Aharonian et al.~2004), HESS (open triangles, Aharonian et al.~2006).
The 5 hr sensitivity of the CTA is marked by the broken dot-dashed line (see Fig.~24 in the paper by The CTA Consortium~2010).}
\label{fig2}
\end{figure}

We calculate the example spectra from the synchrotron process for 
different parameters of the electron spectra which potentially could be responsible for the observed variability of $\gamma$-ray emission reported by the AGILE and the Fermi telescopes. For these same parameters we also calculate the IC $\gamma$-ray spectra in order to show how the end of IC spectrum should vary at energies above 1 TeV. In the calculations we apply the full formulae from Blumenthal \& Gould~(1970). Only the well defined soft photon target have been taken into account when calculating the IC spectra, i.e. the MBR and synchrotron radiation from the nebula. In the case of the synchrotron radio emission, we applied the observations of the Crab Nebula in the GHz energies (Baars \& Hartsuijker~1972). 
The differential photon spectrum of this emission is well  described by a single power law with spectral index 1.26. This synchrotron differential photon density has been obtained by simple averaging over the whole volume of the nebula with the radius of 2 pc. We do not consider other possible low energy radiation fields such as the infrared radiation or the microwave radiation (see. e.g. Atoyan \& Aharonian~1996). The photon densities of these other radiation fields is much more difficult to define preciously in the region of the pulsar wind shock. Moreover their contribution to the total IC $\gamma$-ray spectrum at multi TeV region seems to be less important in respect to the comptonization of the MBR and the radio synchrotron photons (e.g. Atoyan \& Aharonian~1996). 

At first, we model the steady synchrotron spectrum and the flare synchrotron spectrum by changing the maximum energies of accelerated electrons and keeping constant the magnetic field strength at the acceleration region which is fixed on $2\times 10^{-3}$ G. 
For this magnetic field strength electrons have to have Lorentz factors $\gamma_{\rm e} = 7\times 10^8$, in order to be consistent with the observations of the cut-off in the synchrotron spectrum during the quiescent state (see Fig.~2a). 
We also calculate the $\gamma$-ray spectra expected from electrons accelerated to 
maximum energies which might correspond to the emission during the recently observed flare,
applying $\gamma_{\rm br} = 3\times 10^9$, and the supposed super-quiescent stage for which  $\gamma_{\rm br} = 2.3\times 10^8$ (Fig.~2a).
Electrons with the Lorentz factors corresponding to the quiescent stage
scatter the MBR almost in the Klein-Nishina (KN) regime since the energy of the MBR from the peak of the Planck spectrum
is equal to $\varepsilon^{\rm MBR}_{\rm EF} = 3kT\gamma_{\rm e}\approx m_{\rm e}$.
On the other hand, scattering of synchrotron radio emission is regulated to the 
border between T-KN regimes, due to the power law spectrum of the synchrotron photons with the 
spectral index $1.26$. The synchrotron and IC spectra are calculated for the two spectral indexes of electron spectrum, 3.0 and 3.6 (see Fig.~2a,b, respectively). 
As expected the $\gamma$-ray emission produced by these electrons in the GeV and TeV energy ranges show clear correlations. However,
the end of the TeV $\gamma$-ray spectrum from scattering of MBR and synchrotron radio  emission vary on a lower level than the end of the synchrotron spectrum due to the IC scattering in (or close to) the Klein-Nishina regime. The change in the TeV $\gamma$-ray spectrum between the quiescent and flare states is substantial at energies above $\sim 10$ TeV.
Moreover, the IC component, due to the scattering of the radio emission, is steeper and start to dominate at lower energies than the IC component from scattering of the MBR.
The calculation of the level of the IC component from scattering of radio  emission is less reliable due to the uncertain determination of the photon density of the radio synchrotron photons
(produced in the outer nebula) in the region close to the pulsar shock.

In Fig.~2c, we also show the $\gamma$-ray spectra in the case of a very strong magnetic field at the emission region, equal to $B_{\rm sh} = 4\times 10^{-2}$ G, and a few values of the Lorentz factor of electrons at the break of their spectrum. Such strong magnetic field might appear as a result of the compression due to the decelerating pulsar wind.
The Lorentz factors of electrons, producing radiation in the quiescent stage, should be of the order of $\gamma_{\rm br} = 2\times 10^8$. Then, the energies of the MBR photons at the peak of the spectrum are $\varepsilon^{\rm MBR}_{\rm EF}\approx 0.3m_{\rm e}$, i.e. the ICS of the MBR occurs still in the T regime.
Therefore, the cut-offs in the IC spectrum calculated for the electron Lorentz factors corresponding to the quiescent and flaring stages clearly show stronger variability.

\section{Discussion and Conclusion}

We propose that the variable emission from the Crab nebula on a time scale of a few days, recently observed by the AGILE and the Fermi telescopes above 100 MeV, can be understood assuming that the emission region is moving mildly relativistically to the observer.
We suggest that this $\gamma$-ray emission comes from the region just behind the shock in the pulsar wind. It could be produced by electrons accelerated to 
different maximum energies in the electric fields induced during the reconnection process of the magnetic field under the pressure exerted by the shocked wind.
These highest energy electrons produce synchrotron radiation in the MeV-GeV energy range and also the IC $\gamma$-rays above $\sim 1$ TeV by scattering of the MBR and the low energy synchrotron radiation. The relative role of these two soft radiation fields in the ICS process is difficult to estimate due to the lack of precise knowledge on the density of the GHz radio
photons at the acceleration site in the pulsar wind.

We consider different parameters describing the spectrum of injected electrons and compare them with the observations in the MeV-GeV and above 1 TeV energy ranges. 
The flaring stage might result either due to the change of the maximum energies of accelerated electrons (described by their Lorentz factors at the break of the power law spectrum) or the the change of the spectral index of electrons at the highest energies, or by both effects. The change of only the break energy in the electron spectrum results in the TeV $\gamma$-ray IC spectra which show rather low level of variability at energies above $\sim 10$ TeV. Therefore, in such a case it will be difficult to observe synchronous
variations at the GeV and TeV energies even with the future Cherenkov Telescope Array
(see Fig.~2). 
The situation is more promising in the case of the instantaneous change of the break energy in the electron spectrum with the flattening of the electron spectrum. In such a case, the clear variability of the IC spectrum through the broad range of the TeV energies 
could be detected. The possible discovery of flaring emission in the TeV energies will be
easier in the case of significantly stronger magnetic field at the acceleration region. Then, the maximum energies of accelerated electrons can be lower and the IC scattering of the MBR could still occur in the Thomson regime. As a result, the end of IC spectrum should change quadratically with the change of energy of electrons.

The TeV $\gamma$-ray spectra calculated for the range of discussed parameters are compared with the best measurements of the Crab Nebula spectrum provided by the HEGRA (Aharonian et al.~2004) and the HESS (Aharonian et al.~2006). The HEGRA spectrum extends up to $\sim 80$ TeV showing no evidences of the cut-off. On the other hand, the HESS spectrum is steeper and shows
a cut-off at clearly lower energies. Calculated by us TeV spectra from IC scattering of the MBR and synchrotron radio photons in the quiescent and flaring stages shows better consistency with the spectrum measured by the HEGRA Collaboration (Aharonian et al.~2004). The relatively steep TeV $\gamma$-ray spectrum reported by the HESS Collaboration (Aharonian et al.~2006) might be consistent with our calculations provided that it has been obtained during the stage of a relatively non-efficient acceleration of electrons and relatively strong magnetic field strength in the emission region. The sub-GeV synchrotron $\gamma$-ray emission during such stage should be on the level below the $\gamma$-ray flux reported by the recent measurements
by the Fermi-LAT telescope (Abdo et al.~2010). Note however, that the $\gamma$-ray emission reported by the EGRET telescope a decade ago (e.g. Kuiper et al.~2001) seems to be on the lower level than the present Fermi-LAT measurements. This suggest that the sub-GeV emission from the Crab Nebula can vary not only up but also down in respect to the so called quiescent level
observed by the Fermi.

The possible variability of the end of the TeV $\gamma$-ray emission from the Crab Nebula produced in the ICS process by electrons may be partially hidden in the case of efficient acceleration of hadrons within the nebula. These hadrons, coming from the surface of the Crab pulsar, could be also accelerated in the scenario discussed here. They interact with the matter within the Crab Nebula producing steady emission which is expected above $\sim 10$ TeV, see e.g. Amato et al.~(2003), Bednarek \& Bartosik~(2003), Horns  et al.~(2005).

\section*{Acknowledgments}
We would like to thank Maria Giller for discussion.
This work is supported by the Polish MNiSzW grant N N203 390834 and NCBiR grant ERA-NET-ASPERA/01/10.


\label{lastpage}
\end{document}